\documentclass[a4paper,11pt]{article}
\usepackage{amsmath,amsthm,latexsym,amssymb,amsfonts,epsfig}
\usepackage{fontenc,dsfont,layout}
\usepackage{hyperref} 
\usepackage{psfrag}
\usepackage[cp1250]{inputenc}
\usepackage{graphics}
\usepackage{epstopdf} 
\usepackage{caption}
\usepackage{float}
\usepackage{multirow} 
\usepackage{rotating}
 
\numberwithin{equation}{section}

\DeclareMathAlphabet{\mathpzc}{OT1}{pzc}{m}{it}

\begin{document}

\title{Fine-tuning in GGM and the 126 GeV Higgs particle}
\author{Zygmunt Lalak\footnote{Zygmunt.Lalak@fuw.edu.pl} \ and Marek
Lewicki\footnote{Marek.Lewicki@fuw.edu.pl} \\ Institute of
Theoretical Physics, Faculty of Physics, University of Warsaw\\ ul. Ho\.za 69,
Warsaw, Poland} 
\date{}
\maketitle

\begin{abstract}
In this paper we reanalyze the issue of fine-tuning in supersymmetric models which feature Generalized
Gauge Mediation (GGM)  in the light of recent measurement of the mass of the light Higgs particle 
and taking into account available data on the value of the muon magnetic moment $g_\mu-2$. 
We consider GGM models with 3, 5 and 6 input parameters and reduce the fine-tuning by assuming 
simple relations between them at the high scale.  We are able to find solutions which give the 
correct value of the light Higgs mass and are less fine-tuned than models with standard 
gauge mediation (and with gravity mediation), however one never finds fine-tung measure 
lower than about $10^2$ if one neglects the data on $g_\mu-2$ and and about
four times more if one takes the constraint given by $g_\mu-2$ into account. In
general the current $g_\mu-2$ data push the models towards the high fine-tuning region. 
It is interesting to note, that once one 
removes the contributions to the finetuning induced by $\mu$  and $B_\mu$, then in the case with 
neglected $g_\mu-2$ constraint one can easily find realistic vacua with fine-tuning of order $1$ or 
lower, while  the fine-tung remains always large when the $g_\mu-2$ constraint is enforced. One 
should note, that in the last case even a small shift of the light Higgs mass towards smaller values 
both reduces fine-tuning and helps to improve agreement of a model with $g_\mu-2$ data. 
 
\end{abstract}

\section{Introduction} \label{sec:wstep}
 
The discovery of the Higgs boson at LHC with mass of about 126GeV seems to favour
MSSM which predicts that the lightest Higgs boson can't be much heavier than Z
boson. However Higgs mass this far from Z mass requires large radiative
corrections which have to come from heavy supersymmetric particles. Such heavy
sparticles reintroduce some fine-tuning in MSSM \cite{FINETUNE}  because
large supersymmetric parameters also have to cancel out to secure electroweak
symmetry breaking at the right energy scale, thus threatening the motivation of SUSY as solution to
naturalness problem.

In MSSM large fine-tuning originates from requiring sparticles heavy enough
to generate observed Higgs mass. In mSUGRA models the simplest way of
increasing Higgs mass is to get maximal stop mixing which increases
dominant stop correction, but requires large negative A-terms. In gauge mediated
models \cite{GMfirst} however, usually only negligible A-terms are generated at
the SUSY breaking scale. So the Higgs mass can be increased only using non-universality 
of scalars and fermions through subdominant corrections.

General gauge mediation \cite{GGM} has already been studied in terms of
phenomenology \cite{GGMpheno} and specifically fine-tuning \cite{GGMft}.
However, we shall reanalyze the issue of fine-tuning taking into account the
recent measurement of Higgs boson mass and data on the anomalous magnetic moment
of the muon. In this work we use new realization (see the Appendix) of a well known
algorithm used to find SUSY spectra \cite{num}, to check how much fine-tuning can be expected in gauge mediated SUSY breaking models in
the light of recent Higgs boson discovery.

We also redo the same calculation in mSUGRA model using updated experimental
bound on superpartner masses \cite{exp}, and check how calculating fine-tuning
using stability of Higgs mass rather than usual Z mass can improve these
results.

\section{Electroweak breaking in MSSM and fine-tuning}
Neutral part of the scalar potential in MSSM takes the form
\begin{eqnarray}\label{Vhiggssimple}
V&=&(\mu^2+m^2_{H_u})|H_u|^2+(\mu^2+m^2_{H_d})|H_d|^2
+(bH_u H_d+\textrm{h.c})
\nonumber\\
&+&\frac{1}{8}(g^2 +{g'}^2) 
\left( |H_u|^2-|H_d|^2\right)^2.
\end{eqnarray}
Naturalness problem appears in MSSM when we require that the above potential
gives correct electroweak symmetry breaking, which gives us $Z$ boson mass in terms of
superymmetric parameters
\begin{equation}\label{MZeq}                                                     
M_Z^2=
\tan{2\beta}\left(
m_{H_u}^2\tan{\beta}
-m_{H_d}^2\cot{\beta}
\right)-2\mu^2.
\end{equation}
Pushing light Higgs mass to the observed value of $126$GeV requires large
radiative corrections, the biggest one comes from top-stop loop
\cite{Higgsmass}
\begin{equation}\label{higgscorrection}                                                      
\delta m^2_h = \frac{3g^2 m_t^4}{8 \pi^2 m_W^2}
\left[ \log \left( \frac{M_S^2}{m_t^2} \right)
+\frac{X_t^2}{M_S^2} \left( 1-\frac{X_t^2}{12 M_S^2} \right)
\right] ,
\end{equation}
where $M_S^2=\frac{1}{2} \left(m^2_{\bar{t}_1}+m^2_{\bar{t}_2} \right)$ is the
average of stop masses, and $X_t=m_t (A_t-\mu \cot{\beta})$ is an off diagonal
element of stop mass matrix. Parameters in \eqref{MZeq} also receive
top-stop loop corrections
\begin{equation}                                                       
\delta m^2_{H_u} |_{stop} = -\frac{3Y_t^2}{8 \pi^2}
\left( m^2_{Q_3} + m^2_{U_3} + |A_t|^2 \right)
\log\left(\frac{M_u}{\textrm{TeV}}\right),
\end{equation}
where $m^2_{Q_3}$ $m^2_{U_3}$ and $A_t$ are supersymmetric parameters that
predict the stop mass, and $M_u$ is a scale at which soft masses are generated.

So requiring correct Higgs mass gives large corrections that have to cancel out
on the right hand side of \eqref{MZeq} to give the correct $M_Z$.

We define fine-tuning measure with respect to parameter $a$ as
follows \cite{Chankowski}
\begin{equation}                                                        
\Delta_a=\left| \frac{\partial \ln{M^2_Z}}{\partial \ln{a}} \right|.
\end{equation}
Fine-tuning connected with a set of independent parameters $a_i$ is then
\begin{equation}\label{Delta}                                                      
\Delta=\max_{a_i} \Delta_{a_i}.  
\end{equation}
Remembering that fine-tuning in the Standard Model actually appeared in the
Higgs boson mass we can define fine-tuning with respect to Higgs mass in MSSM 
\begin{equation}                                                        
\Delta_{h \ a}=\left| \frac{\partial \ln{m^2_h}}{\partial \ln{a}} \right| \ \ 
; \ \ 
\Delta_h=\max_{a_i} \Delta_{h \ a_i},
\end{equation}
and calculate it numerically similarly to fine-tuning with respect to $Z$ mass.
In our Figures we plot several regions of allowed solutions corresponding to
different models on top of each other, because only the borders of these regions
(corresponding to the minimal fine-tuning) are actually important.
\begin{figure}[ht]
\centering    
\input{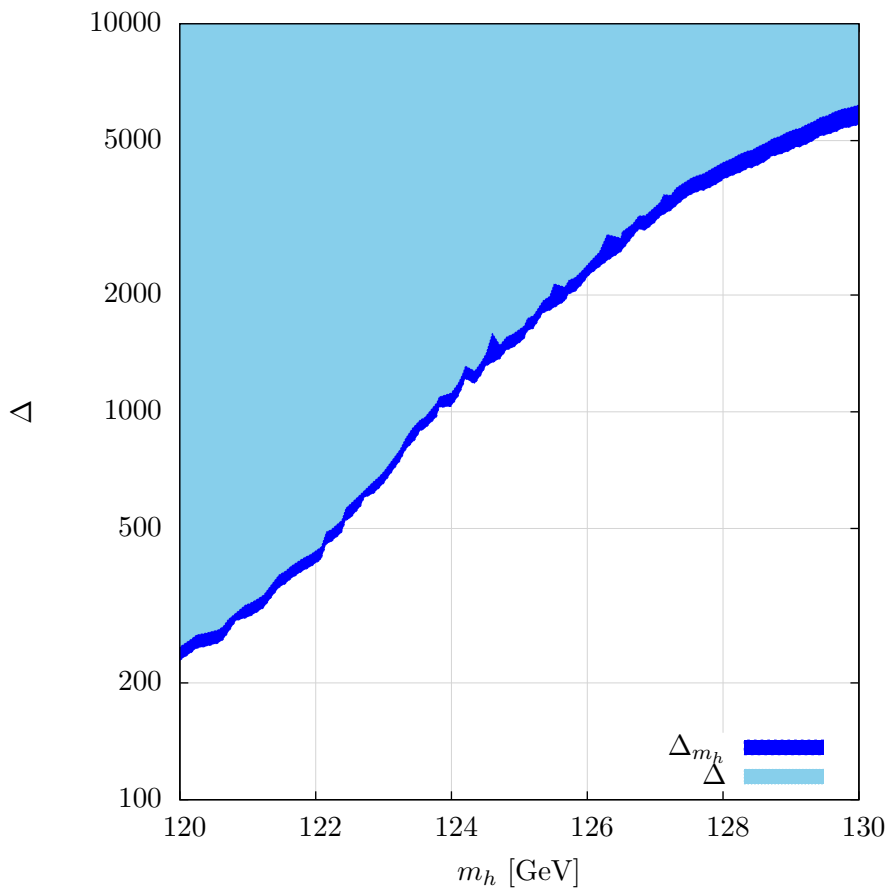} 
\parbox{\textwidth}{
\captionof{figure}{ 
Fine-tuning from Higgs mass and $Z$ mass in mSUGRA model with
soft terms generated at scale $M_u=2.5\times10^{16}\textrm{GeV}$ and with
$\tan{\beta}=40$.\label{ftmsugra} }
}  
  
\end{figure} 
Figure~\ref{ftmsugra} shows that as expected fine-tuning from Higgs boson mass
turns out to be similar to the one obtained from $Z$ boson mass and usually is a few percent lower.
\section{Gravity vs Gauge mediation}
Meade, Shih and Seiberg \cite{GGM} defined gauge mediated models as those in
which visible and hidden sectors decouple when gauge couplings vanish. They also
have shown that in general such models can only have six parameters determining
the low energy sparticle spectrum. In this work we parametrise the high energy soft
SUSY breaking terms with three parameters corresponding to gaugino masses
$M_1=m_Y$ , $M_2=m_w$ , $M_3=m_c$ and three parameters determining scalar masses
$\Lambda^2_c$ , $\Lambda^2_w$ , $\Lambda^2_Y$ which give
\begin{equation}\label{sfermionmasses}                                                  
m^2_{f}=2 \left[ 
C_3(f)\left(\frac{\alpha_3}{4 \pi}\right)^2\Lambda^2_c +
C_2(f)\left(\frac{\alpha_2}{4 \pi}\right)^2\Lambda^2_w +
C_1(f)\left(\frac{\alpha_1}{4 \pi}\right)^2\Lambda^2_Y
 \right],
\end{equation}
where $\alpha_i = g^2_i/4 \pi^2$ and
\begin{eqnarray}
C_1(f)&=&\frac{3}{5}Y^2_f  \nonumber \\
C_2(f) &=& \left\{ 
\begin{array}{l l}
  \frac{3}{4} & \quad \mbox{for} \quad f = Q , L , H_u , H_d \\
  0 & \quad \mbox{for} \quad f = U , D , E \\
\end{array} \right.
\\ 
C_3(f) &=& \left\{  
\begin{array}{l l}
  \frac{4}{3} & \quad \mbox{for} \quad f = Q , U , D \\
  0 & \quad \mbox{for} \quad f = E , L , H_u , H_d. \\
\end{array} \right.  \nonumber
\end{eqnarray} 
A specific model of gauge mediation gives above quantities in terms of physical
parameters present in the model. As an example we use two of the models
published in \cite{GGMmodels}, the first of which (GGM1) is defined by the
superpotential
\begin{equation}                                                     
W_1 = X_i ( y^i \bar{Q}Q + r^i \bar{U}U + s^i \bar{E}E ),
\end{equation}
with three independent parameters used to calculate soft masses 
\begin{equation} 
\Lambda_Q=\frac{y^i F_i}{y^j X_j} \quad
\Lambda_U=\frac{r^i F_i}{r^j X_j} \quad
\Lambda_E=\frac{s^i F_i}{s^j X_j}.
\end{equation}
In terms of which  soft masses take the form
\begin{eqnarray}                                                     
m_c = \frac{\alpha_3}{4 \pi}(2\Lambda_Q+\Lambda_U ),\quad \quad
m_w &=& \frac{\alpha_2}{4 \pi}3\Lambda_Q, \quad \quad
m_Y = \frac{\alpha_1}{4 \pi}\left(
\frac{4}{3}\Lambda_Q + \frac{8}{3}\Lambda_U + 2\Lambda_E
\right), \nonumber \\   
\Lambda^2_c = 2\Lambda^2_Q
+\Lambda^2_U, \quad \quad
\Lambda^2_w &=& 3\Lambda^2_Q, \quad \quad
\Lambda^2_Y =
\frac{4}{3}\Lambda^2_Q+\frac{8}{3}\Lambda^2_Q+2\Lambda^2_E. \\
\end{eqnarray}
The second model (GGM2) is defined by
\begin{equation}                                                     
W_2 = X_i ( y^i \bar{Q}Q + r^i \bar{U}U + s^i \bar{E}E+  \lambda^i_q q \tilde{q}
+ \lambda^i_l l \tilde{l})+ F^i X_i,
\end{equation}
with five independent parameters used to calculate soft masses 
\begin{equation} 
\Lambda_Q=\frac{y^i F_i}{y^j X_j} \quad
\Lambda_U=\frac{r^i F_i}{r^j X_j} \quad
\Lambda_E=\frac{s^i F_i}{s^j X_j} \quad
\Lambda_q=\frac{\lambda_q^i F_i}{\lambda_q^j X_j} \quad
\Lambda_l=\frac{\lambda_l^i F_i}{\lambda_l^j X_j}.
\end{equation}
Again we obtain soft masses of the form
\begin{eqnarray}                                                     
m_c &=& \frac{\alpha_3}{4 \pi}(\Lambda_q+2\Lambda_Q+\Lambda_U ),\quad  \quad
m_w = \frac{\alpha_2}{4 \pi}(\Lambda_l+3\Lambda_Q), \nonumber \\
m_Y &=& \frac{\alpha_1}{4 \pi}\left(
\frac{2}{3}\Lambda_q  + \Lambda_l +
\frac{4}{3}\Lambda_Q + \frac{8}{3}\Lambda_U + 2\Lambda_E
\right),  \\   
\Lambda^2_c &=& \Lambda^2_q+2\Lambda^2_Q+\Lambda^2_U, \quad  \quad
\Lambda^2_w = \Lambda^2_l+3\Lambda^2_Q, \quad \quad
\Lambda^2_Y =\frac{2}{3}\Lambda^2_q+\Lambda^2_l+
\frac{4}{3}\Lambda^2_Q+\frac{8}{3}\Lambda^2_U+2\Lambda^2_E. \nonumber
\end{eqnarray}

Main disadvantage of gauge mediation in respect to of fine-tuning comes from
the fact that only negligible $A$-terms are generated at SUSY breaking scale. Large
mixing in the sfermion mass matrices would increase its contribution to Higgs
mass as in eq.\eqref{higgscorrection}, and make it easier to achieve the
experimental result of Higgs boson mass. 
On the other hand prediction of
nonuniversal gaugino masses makes it easier to avoid experimental bound
on gluino mass. Nonuniversal scalar masses help avoiding bounds on masses of the
first and second generation squraks. We use the following bounds on
sparticle masses \cite{exp}
\begin{eqnarray}\
m_g & \geq & 1500 \textrm{GeV}, \nonumber \\
m_{u_i},m_{d_i},m_{c_i},m_{s_i} & \geq & 1500 \textrm{GeV} \quad i=1,2,
\nonumber \\ m_{t_i} & \geq & 560 \textrm{GeV} \quad i=1,2, \\
m_{b_i} & \geq & 620 \textrm{GeV} \quad i=1,2, \nonumber \\
m_{\chi_1} & \geq & 250 \textrm{GeV}.  \nonumber 
\end{eqnarray}
We also assume $M_u=10^{8}\textrm{GeV}$ and $\tan{\beta}=40$.
\begin{figure}[ht] 
\centering     
\input{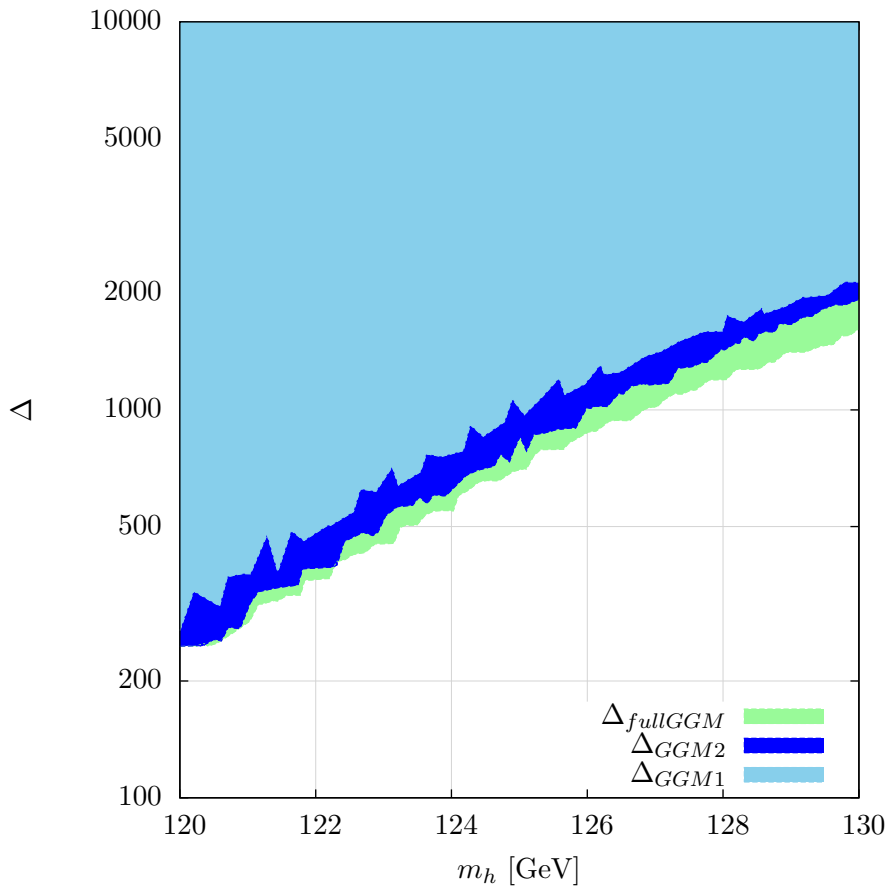}  
\parbox{\textwidth}{
\captionof{figure}{
Fine-tuning in models GGM1 and GGM2 as well as in the general six parameter
case. \label{ggmft} 
} 
}   
\end{figure}
Figure~\ref{ggmft} shows that generally models with larger number of free
parameters predict smaller fine-tuning because they allow to increase Higgs
boson mass with subdominant corrections.

In a general model with $6$ parameters, the biggest sources of fine-tuning are
the gluon mass parameter $m_c$ or contributions to scalar masses connected with
color $\Lambda^2_c$ or weak interactions $\Lambda^2_w$. The $\mu$ parameter
contribution is small in solutions that minimize fine-tuning for a given Higgs mass, because $\mu$ can be decreased by increasing
$\Lambda^2_Y$ and $\Lambda^2_w$ and decreasing $\Lambda^2_c$ which increases
high scale $m^2_{H_u}$ while keeping masses of coloured particles fixed. The
walue of $\mu$ decreases with decreasing energy scale and eventually runs
negative to secure right elektroweak symmetry breaking, as we can see from large
$\tan \beta$ approximation of \eqref{MZeq}
\begin{equation}
\frac{m^2_Z}{2} \approx -m_{H^2_u}-|\mu|^2. 
\end{equation}
As we can see increasing high scale $m^2_{H_u}$ makes it run down towards
smaller negative value and so decreases $\mu$ required to obtain correct Z mass.
Since colored particle masses that would increase overall fine-tuning aren't
changed we obtain a scenario with smaller $\mu$ parameter and similar
fine-tuning. Meanwhile, increased $\Lambda^2_Y$ and $\Lambda^2_w$ give us larger
sub dominant corrections to Higgs mass due to increased masses of non coloured
particles.

In model GGM2 squark and gluino masses obtain contributions from all parameters
connected with color interactions $\Lambda_Q , \Lambda_U , \Lambda_q$ and
fine-tuning coming from these masses is distributed among these fundamental
parameters. The largest fine-tuning contribution turns out to come typically 
from the $\mu$ parameter and can come from one of the parameters connected with
color only if said parameter is much larger than the other two.

The same can be said about the model GGM1. The biggest source of fine-tuning
is usually $\mu$, except cases where one of the other parameters is much larger
than the other two.

\section{Reduction of fine-tuning in GGM}
The simplest way of reducing fine-tuning is assuming we are considering a model
that predicts the parameters which are not independet of each other, but instead
are functions of some fundamental parameters. For example, if gaugino
masses $M_i$ are given functions of parameter $M_{\frac{1}{2}}$ we obtain
\begin{eqnarray}
 M_i &=& f_i(M_{\frac{1}{2}}), \nonumber \\ 
\frac{1}{2}\Delta_{M}
&=&
\left| \frac{\partial \ln{M_Z}}{\partial \ln{M_{\frac{1}{2}}}} \right|
= \left| \frac{M_{\frac{1}{2}}}{M_z}
\frac{\partial M_Z}{\partial M_{\frac{1}{2}}} \right|
= \left| \frac{M_{\frac{1}{2}}}{M_z}
\frac{\partial M_Z}{\partial M_i}
\frac{\partial M_i}{\partial M_{\frac{1}{2}}} \right| \\
&=& \left| \frac{M_{\frac{1}{2}}}{M_z}
\frac{M_i}{M_i}f'_i(M_{\frac{1}{2}})
\frac{\partial M_Z}{\partial M_i} \right|
= \left| M_{\frac{1}{2}} \frac{f'_i(M_{\frac{1}{2}})}{f_i(M_{\frac{1}{2}})}
\frac{M_i}{M_Z}
\frac{\partial M_Z}{\partial M_i} \right| \nonumber \\
&=& \left|
M_{\frac{1}{2}} \frac{f'_i(M_{\frac{1}{2}})}{f_i(M_{\frac{1}{2}})}
\frac{\partial \ln{M_Z}}{\partial \ln{M_i}}
\right|
=\left| \sum\limits_{i=1}^3
c_i(M_{\frac{1}{2}}) \frac{\partial \ln{M_Z}}{\partial \ln{M_i}}
\right|
. \nonumber 
\end{eqnarray}
If $f_i$ are simply proportional to $M_{\frac{1}{2}}$ one finds
\begin{equation}
\Delta_{M}=
\left| \sum\limits_{i=1}^3 
\frac{\partial \ln{M^2_Z}}{\partial \ln{M_{i}}}
\right|. \nonumber
\end{equation}
If these functions were logarithms  
\begin{eqnarray}
 f_i(M_{\frac{1}{2}}) &=& \tilde m \ln{\frac{M_{\frac{1}{2}}}{Q}}, \\
c_i(M_{\frac{1}{2}}) &=& \frac{\tilde m}{M_i} .
\nonumber 
\end{eqnarray}
Keeping in mind that fine-tuning is proportional to soft terms 
$\Delta_{M_i} \propto M_i$, we obtain $\Delta_{M_{\frac{1}{2}}} \propto \tilde
m$. \\ We now check how the usual proportionality of nonuniversal soft masses and
$\mu$ parameter helps reducing fine-tuning in models GGM1 and GGM2 as well as in
the general case.
\begin{figure}[ht]
\begin{minipage}[t]{0.45\linewidth} 
\centering
\include{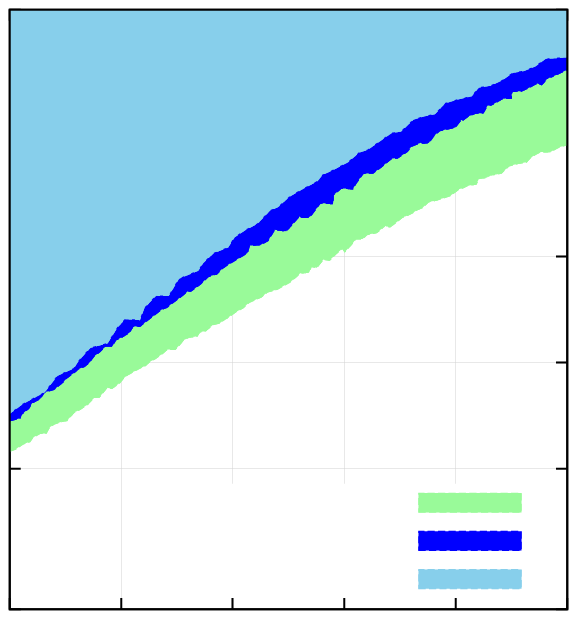} 
\end{minipage}
\hspace{0.5cm}
\begin{minipage}[t]{0.45\linewidth}
\centering 
\include{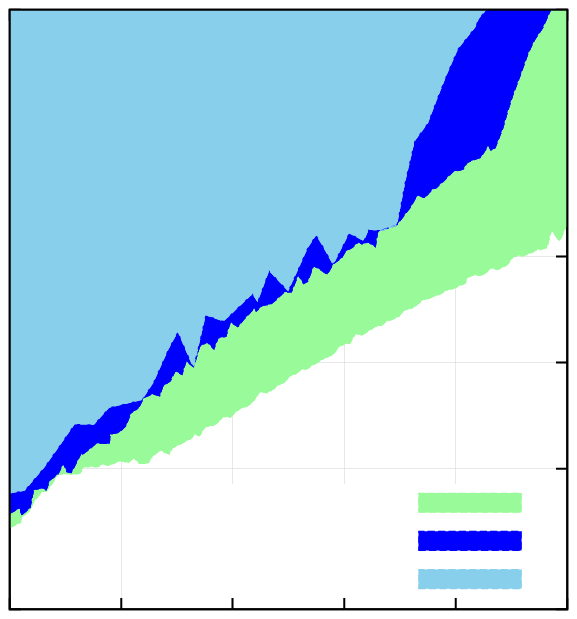}  
\end{minipage}
\captionof{figure}{
Fine-tuning in models GGM1 and GGM2 as well as in the general six parameter
case with parameters defining superparticle spectrum proportional to each other
and to the $\mu$ parameter.\label{ggmftn}
}
\end{figure}
As one can see from Figure~\ref{ggmftn},
 simple proportionality of soft terms can greatly decrease fine-tuning in GGM
 but one still finds $\Delta>100$ for $m_h=126$, even in the most general case.
\begin{figure}[ht] 
\centering    
\input{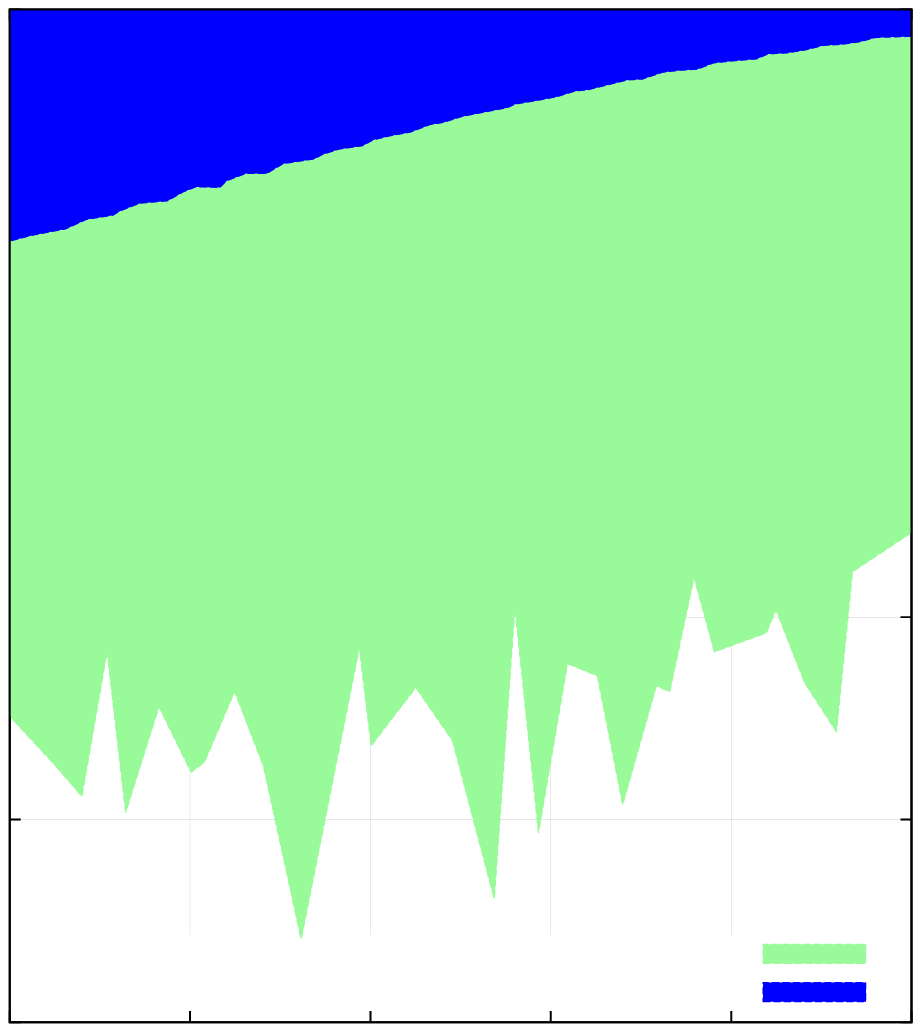}  
\parbox{\textwidth}{
\captionof{figure}{ 
Fine-tuning in model GGM1 with and without contribution to fine-tuning from
$\mu$ and $B_\mu$ parameters and with soft terms proportional to each other.
\label{ggm2nomu} }  
}  
\end{figure}
We have also checked that in models considered here (for example GGM1 in
Figure~\ref{ggm2nomu}) fine tuning coming only from the gauge mediated soft
terms can cancel out very precisely if they are proportional to one another, as
pointed out in \cite{Brummer}. 
\section{Constraints from $g_\mu - 2$}  
In this section we check whether discussed models can accommodate the
discrepancy between measured muon magnetic moment and the standard model prediction
\cite{damu}   
\begin{equation}
\delta a_\mu = a_{\mu}^{\textrm{EXP}}-a_{\mu}^{\textrm{SM}}=(2.8 \pm
0.8)10^{-9} .
\end{equation}
The simplest approximation of supersymmetric contribution to muon magnetic   
moment is obtained by assuming that $\tan\beta$ is large and all masses in
slepton sector are equall to $M_{\textrm{SUSY}}$. This way one obtains
\cite{g-2}
\begin{equation}
\delta a_\mu^{\textrm{SUSY}} \approx \left(  \frac{g_1^2-g_2^2}{192 \pi^2}
+\frac{g_2^2}{32 \pi^2} \right) \frac{m_{\mu}^2}{M_{\textrm{SUSY}}^2}\tan\beta,
\end{equation} 
which indicates a problem since Higgs boson mass depends on soft breaking terms
only logarithmically.
We evaluate $\delta a_\mu^{\textrm{SUSY}}$ numerically using full 1-loop 
SUSY corrections and 2-loop QED logarithmic corrections from \cite{g-2}. 
\begin{figure}[ht]
\begin{minipage}[t]{0.45\linewidth}  
\centering
\include{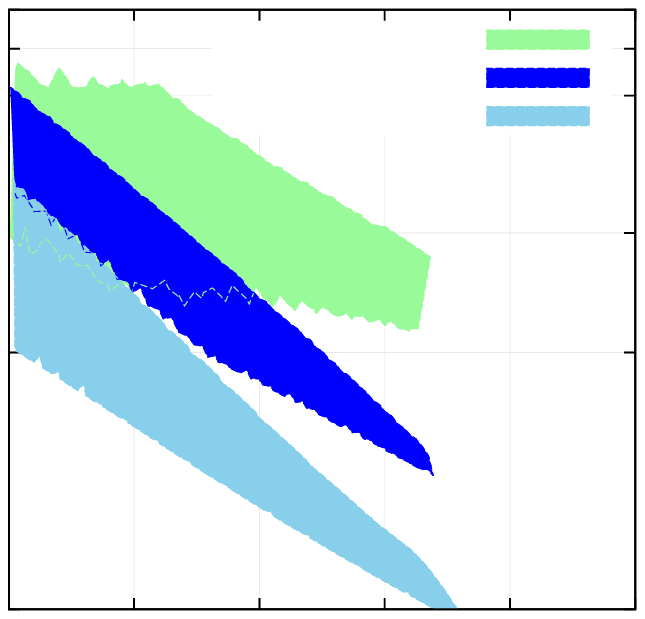}
\end{minipage}
\hspace{0.5cm}   
\begin{minipage}[t]{0.45\linewidth}
\centering
\include{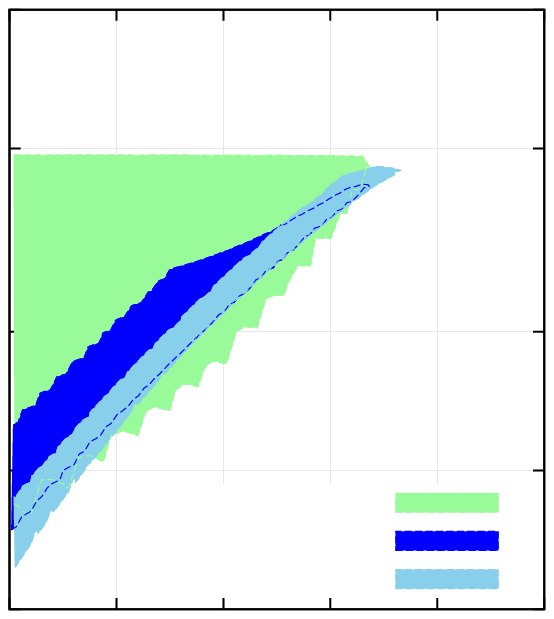}
\end{minipage}
\captionof{figure}{  
Regions of largest possible SUSY contribution to muon g-2 and corresponding
fine-tuning \label{ggmdamu}}
\end{figure}
From Figure~\ref{ggmft} we can see that only the general case predicts $\delta
a_{\mu}$ within $1\sigma$ bound for $m_h=126$, while other models fall out of
$2\sigma$ bounds. Even in the most general case it is hard to increase $\delta
a_{\mu}$ because all slepton generations have the same mass at the high scale.
The 3rd generation gets negative contribution from large Yukawa coupling
\begin{equation}
16 \pi^2 \frac{d}{dt}m^2_{L_3} \supset
2|h_{\tau}|^2(m^2_{H_d}+m^2_{L_3}+m^2_{E_3}+A_{\tau}^2)
\end{equation}
which can make stau tachionic before smuon is light enough to produce the
required value of  $\delta a_{\mu}$.

Also requiering small masses in slepton sector means we can only
increase Higgs mass with dominant squark corrections which increase fine-tuning.
And we are left only with solutions with much higher fine-tuning than those
that use sub dominant corrections to Higgs mass which we described in previous
chapters.
\section{Summary and conclusions} 
We have reanalyzed the issue of fine-tuning in supersymmetric models which feature Generalized 
Gauge Mediation (GGM)  in the light of recent discovery of the $126$ GeV  Higgs particle and 
taking into account available data on the value of the muon magnetic moment $g_\mu-2$. 
We consider GGM models with 3, 5 and 6 input parameters and reduce the
fine-tuning by assuming simple relations between them at the high scale.  We are able to
find solutions which give the correct value of the light Higgs mass and are
less fine-tuned than models with standard gauge mediation, however one never
obtains fine-tung measure lower than about $10^2$ if one neglects the data on
$g_\mu-2$ and and about four times more if one takes the constraint given by
$g_\mu-2$ into account. In general the current $g_\mu-2$ data push the models
towards high fine-tuning region. However, it is interesting to study the
fine-tuning after removing the contributions to the fine-tuning induced by
$\mu$  and $B_\mu$, since it isn't obvious that the origin of these two
parameters has anything to do with gauge mediation. It is interesting to note,
that once this is done, then in the case with neglected $g_\mu-2$ constraint
one can easily find realistic vacua with purely gauge mediated fine-tuning of
order $1$ or lower, while  the fine-tung remains always large when the $g_\mu-2$ constraint is enforced. One should note, that in the last case even a small shift of the light Higgs mass towards smaller values both reduces fine-tuning and helps to improve agreement of a model with $g_\mu-2$ data. Decrease of the Higgs mass down to $123$ GeV reduces the fine-tuning by a factor of $2$.

To sum up, in models featuring GGM one can naturally obtain fine-tuning smaller
than that in models with gravity mediation, despite vanishing A-terms at the
high scale. Moreover, considering exclusively fine-tuning coming from
gauge-mediated soft masses one can easily achieve arbitrarily small fine-tuning
while staying with the correct value of the light Higgs mass. Imposing the
agreement of the model with the $g_\mu - 2$ data restricts parameter space to
the region of enlarged fine-tuning, but it is possible to find models which fit
into the $1\sigma$ band. Even a small decrease of the measured value of the Higgs mass would allow for much better agreement of GGM models with measured $g_\mu-2$.
\begin{center}
{\bf Acknowledgements}
\end{center}
 Authors thank S. Pokorski for very helpful discussions.
 ZL thanks D. Ghilencea for discussions. \\ 
This work was supported by the Foundation for
Polish Science International PhD Projects Programme co-financed by the EU
European Regional Development Fund. This work has been partially supported by Polish Ministry for Science and Education
under grant N N202 091839, by National Science Centre under research grant DEC-2011/01/M/ST2/02466 and by National Science Centre under research grant DEC-2011/01/M/ST2/02466.

\appendix 
\section{Numerical procedure}
The numerical procedure we used is similar to the ones used in existing codes
like \cite{num}. We work with quantities
renormalized in $\overline{DR}$ and use renormalization group equations (RGE),
to iteratively find low energy parameters for a given set of high energy set
terms.

\begin{figure}[ht]
\setlength{\unitlength}{0.25cm} 
\centering 
\begin{picture}(40,50)
\put(1,35){\framebox(35,10){
\parbox{33\unitlength}{
calculate radiative corrections to couplings 
$g_i(M_Z)$,$h_t(M_Z)$,$h_b(M_Z)$,$h_\tau(M_Z)$
(use SM values in the first run)
}
}}
\put(15,35){\vector(0,-1){5}}\put(16,32) {$\textrm{RGE}:M_z \rightarrow M_u$}
\put(1,25){\framebox(35,5){
\parbox{33\unitlength}{
include soft breaking terms given at high scale $M_u$
}
}}
\put(15,25){\vector(0,-1){5}}\put(16,22) {$\textrm{RGE}:M_u \rightarrow
M_{EWSB}$}
\put(1,10){\framebox(35,10){
\parbox{33\unitlength}{
iteratively calculate $\mu$,$B_\mu$ and the mass spectrum
(in the first run find estimates for $M_{EWSB}$
,$\mu$ and $B_\mu$)
}
}}
\put(15,10){\vector(0,-1){5}}\put(16,7) {if $\mu$ converged}
\put(1,0){\framebox(35,5){
\parbox{33\unitlength}{
calculate physical masses
}
}}
\put(36,15){\line(1,0){2}}\put(36,40){\line(1,0){2}}
\put(38,15){\vector(0,1){10}}\put(38,15){\line(0,1){25}}
\put(39,16) {
\begin{sideways}
$\textrm{RGE}:M_{EWSB} \rightarrow M_Z$
\end{sideways}
}
\end{picture}
\parbox{40\unitlength}{
\caption{
Schematic of the algorithm we used. all the steps are described in the appendix.
} 
}
\label{algorytm}
\end{figure}
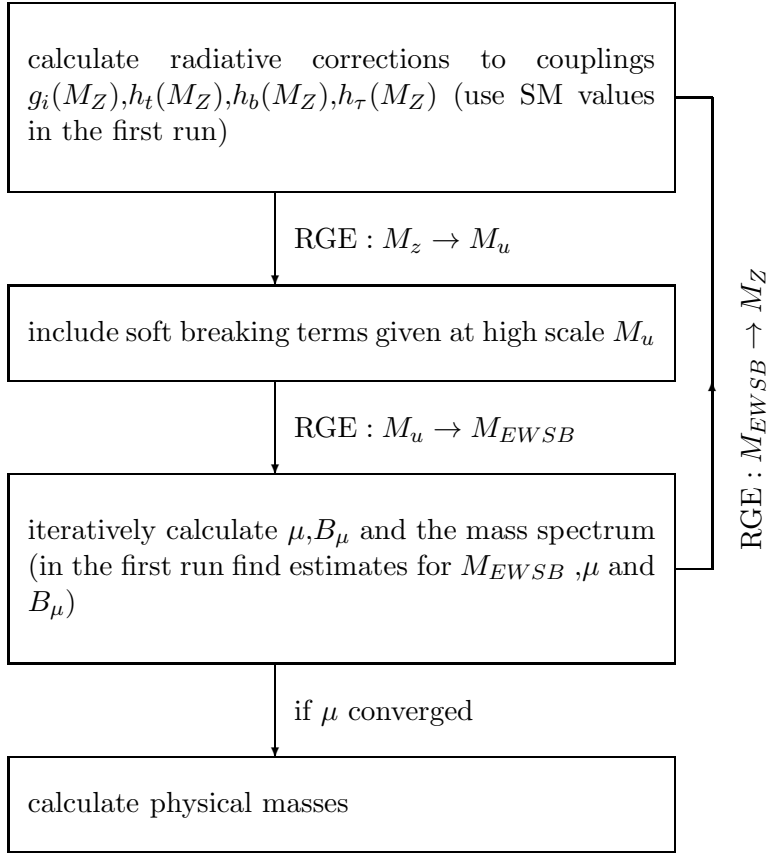
\subsection{$M_Z$ Scale}
At $M_Z$ scale we include radiative corrections to couplings. We set Yukawa
couplings using tree level relations
\begin{equation}
h_t = \frac{m_t \sqrt{2}}{v \sin{\beta}} \quad , \quad
h_b = \frac{m_b \sqrt{2}}{v \cos{\beta}} \quad , \quad
h_{\tau} = \frac{m_{\tau} \sqrt{2}}{v \cos{\beta}},
\end{equation}
where $m_t , m_b , m_{\tau}$ are fermion masses and $v$ is the Higgs field
void expectation value. At first iteration we use physical masses and SM Higgs
vev $v\approx 246,22$. At next iterations above quantities are renormalized in
$\overline{DR}$ scheme and one-loop corrections are included.
To calculate top mass we use 2-loop QCD corrections \cite{Avdeev}
and 1-loop corrections from super partners from the appendix of
\cite{BPMZ}. While calculating bottom mass we follow \textit{Les Houches Accord} 
\cite{accord}, starting from running mass in $\overline{MS}$ scheme in SM
${m_b}^{\overline{MS}}_{SM}$. Next applying procedure described in \cite{MbDR}
we find  $\overline{DR}$ mass at $M_Z$, from which we get MSSM value by
including corrections described in appendix D of \cite{BPMZ}.
While calculating tau mass we include only leading corrections from \cite{BPMZ}.
We calculate Higgs vev in MSSM using
\begin{equation}
v^2=4 \frac{M^2_Z+\Re{\Pi^T_{ZZ}(M_Z)}}{g^2_2+3 g^2_1 /5},
\end{equation}
where we include $Z$ self interactions described in appendix D of \cite{BPMZ}.
To calculate $g_1$ , $g_2$ i $g_3$ in $\overline{DR}$ in MSSM we use procedure
described in appendix C of \cite{BPMZ}.
\subsection{RGE and $M_u$ scale}
after calculating coupling constants at $M_Z$ scale
we numerically solve renormalization group equations 
\cite{Martin},\cite{Yamada}, to find their values at $M_u$ scale, at which we 
include the soft breaking terms. Then we solve RGEs again to get sot terms,
coupling constants, $\tan{\beta}$ and Higgs vev
$v$ at scale
$M_{EWSB}=\sqrt{m_{\bar{t}_1}(M_{EWSB})m_{\bar{t}_2}(M_{EWSB}})$. 
At first iteration we take  $\mu=\textrm{sgn}(\mu)1\textrm{GeV}$
and $B_\mu=0$ and run to scale at which the above equation is fulfilled.Next
using potential minimization conditions we find new values of $\mu$ and $B_\mu$.
\subsection{Calculation of physical masses}
To calculate physical masses we use only leading corrections described in
\cite{BPMZ} everywhere but the Higgs sector.
In the Higgs masses calculation we use full one-loop corrections from
\cite{BPMZ} and leading two-loop corrections described in \cite{Slavich}. 
\subsection{Constraints imposed on the scalar potential}
To chceck if a given set of soft terms describes a realistic physical situation
we check if the scalar potential is not unbounded from below (UFB). And if the
potential dose not have minimums deeper than the one breaking electro-weak
symmetry, which would break $SU(3)$ or $U(1)_{em}$ (CCB).
\cite{UFBCCB}. We include simple tree level bounds:
\begin{itemize} 
\item for UFB
\begin{equation}\label{UFB}
|\mu B_\mu| \le m^2_{H_u}+m^2_{H_d} \quad  
\textrm{at scale} \quad M_x \in [M_{EWSB},M_u],
\end{equation}
\item and CCB
\begin{equation}\label{CCB}
A^2_f \le 3( m^2_{f_L}+m^2_{f_R}+\mu^2+m^2_{H_u} ) \quad 
\textrm{at scale} \quad M_x \in [M_{EWSB},M_u].
\end{equation} 
\end{itemize}

\end{document}